\documentclass[a4paper,12pt]{article} 
\usepackage{graphicx} 
\usepackage{graphics} 
\usepackage{amsmath} 
\usepackage{amsfonts}  
\usepackage{amssymb} 

\usepackage[colorlinks=true, pdfstartview=FitV, linkcolor=blue,  citecolor=blue, urlcolor=blue]{hyperref}  

\newcommand{\dfr}{\widehat{d}} 

\newcommand{\Lu}{\mathbb{L}} 
 
\newcommand{\Pl}{\mathbb{P}} 
\newcommand{\Vsc}{{\cal V}} 
\newcommand{\Ds}{{\cal D}} 

\def\l{\left}
\def\r{\right}
\def\be{\begin{equation}}
\def\ee{\end{equation}}
\def\beq{\begin{equation}}
\def\eeq{\end{equation}}
\def\d{\partial}

\begin{document}

\title{Macroscopic quantum-type potentials in scale relativity}
\author{Laurent Nottale\\{\small CNRS, LUTH, Paris Observatory and Paris-Diderot University} \\{\small 92195 Meudon CEDEX, France}\\
{\small laurent.nottale@obspm.fr}}
\maketitle

\begin{abstract}
We review in this paper the use of the theory of scale relativity and fractal space-time as a tool particularly well adapted to the possible development of a future genuine theoretical systems biology. We emphasize in particular the concept of quantum-type potentials, since in many situations the effect of the fractality of space -- or of the underlying medium -- amounts to the addition of such a potential energy to the classical equations of motion. Various equivalent representations -- geodesic, quantum, fluid mechanical, stochastic -- of these equations are given, as well as several forms of generalized quantum potentials. Examples of their possible intervention in high critical temperature superconductivity and in turbulence are also described, since some biological processes may be analog in some aspects to these physical phenomena. These potential energy extra contributions could have emerged in biology from the very fractal nature of the medium, or from an evolutive advantage, since they involve spontaneous properties of self-organization, morphogenesis structuration and multi-scale integration.
 \end{abstract}
 
 \section{Introduction}

The theory of scale relativity and fractal space-time accounts for a possibly nondifferentiable geometry of the space-time continuum, basing itself on an extension of the principle of relativity to scale transformations of the reference system. Its framework revealed to be particularly adapted to a new theoretical approach of systems biology \cite{Auffray2008, Nottale2008, Noble2008}.

This theory has been initially built with the goal of re-founding quantum mechanics on prime principles \cite{Nottale1984, Nottale1989, Nottale1993}. The success of this enterprise \cite{Nottale2007C,Nottale2011} has been completed by obtaining new results: in particular, a generalisation of standard quantum mechanics at high energy to new forms of scale laws \cite{Nottale1992}, and the discovery of the possibility of macroscopic quantum-type behavior under certain conditions \cite{Nottale1997A}, which may well be achieved in living systems.

This new ``macroquantum" mechanics (or ``mesoquantum" at, e.g., the cell scale) no longer rests on the microscopic Planck constant $\hbar$. The parameter which replaces $\hbar$ is specific of the system under consideration, emerges from self-organization of this system and can now be macroscopic or mesoscopic. This theory is specifically adapted to the description of multi-scale systems able to spontaneous self-organization and structuration.  Two priviledged domains of applications are therefore astrophysics  \cite{Nottale1993, Nottale1996, Nottale1997A, Nottale1997C, Nottale2000C, Nottale2011} and biophysics \cite{Nottale2000B, Auffray2008, Nottale2008, Nottale2009B, Nottale2011}.

In this contribution dedicated to applications in  biophysics, after a short reminder of the theory and of its methods and mathematical tools, we develop some aspects which may be relevant to its explicit use in effective biophysical problems. A special emphasis is made of the concept of macroquantum potential energy. Scale relativity methods are relevant because they provide new mathematical tools to deal with scale-dependent fractal systems, like equations in scale space and scale-dependent derivatives in physical space. This approach is also very appropriate for the study of biological systems because its links micro-scale fractal structures with the organized form at the level of an organism.

 For more information the interested reader may consult the two detailed papers \cite{Auffray2008,Nottale2008} and references therein. 

\section{Brief reminder of the theory}

The theory of scale relativity consists of introducing in an explicit way the scale of measurement (or of observation) $\varepsilon$ in the (bio-)physical description. These scale variables  can be identified, in a theoretical framework, to the differential elements $\varepsilon=dX$, and, in an experimental or observational framework, to the resolution of the measurement apparatus. 

The coordinates can now be explicit functions of these variables, $X=X(dX)$ (we omit the indices for simplicity of the wrtiting, but the coordinates are in general vectors while the resolution variables are tensors \cite[Chap. 3.6]{Nottale2011}). In case of divergence of these functions toward small scales, they are fractal coordinates. The various quantities which describe the system under consideration become themselves fractal functions,  $F=F[X(dX),dX]$. In the simplified case when the fractality of the system is but a consequence of that of space, there is no proper dependence of $F$ in function of $dX$, and we have merely $F=F[X(dX)]$.

The description of such an explicitly scale dependent system needs three levels instead of two.  Usually, one makes a transformation of coordinates $X \to X + dX$, then one looks for the effect of this infinitesimal transformation on the system properties, $F \to F + dF$. This leads to write differential equations in terms of space-time coordinates. 

But in the new situation, since the coordinates are now scale dependent, one should first state the laws of scale transformation, $\varepsilon \to \varepsilon'$, then their consequences on the coordinates, $X(\varepsilon) \to X'(\varepsilon')$ and finally on the various (bio-)physical quantities $F[X(\varepsilon)]  \to F'[X'(\varepsilon')]$.  One of the main methods of the scale relativity theory just consists of describing these scale transformations by differential equations playing in scale space (i.e., the space of the scale variables $\{\varepsilon\}$). In other words, one considers infinitesimal scale transformations, $\ln(\varepsilon/ \lambda) \to \ln(\varepsilon/ \lambda)+ d  \ln(\varepsilon/ \lambda)$, rather than the discrete iterated transformations that have been most often used in the study of fractal objects \cite{Mandelbrot1975, Mandelbrot1982, Barnsley1988}. 

The motion equations in scale relativity are therefore obtained in the framework of a double partial differential calculus acting both in space-time (positions and instants) and in scale space (resolutions), basing oneself on the constraints imposed by the double principle of relativity, of motion and of scale.

\subsection{Laws of scale transformation}
The simplest possible scale differential equation which determines the length of a fractal curve (i.e., a fractal coordinate) ${\cal L}$ reads
\beq
\frac{\d {\cal L} }{\d \ln \varepsilon}=a+b{\cal L},
\label{eq.4}
\eeq
where $\d /\d \ln \varepsilon$ is the dilation operator \cite{Nottale1992, Nottale2011}. Its solution combines a self-similar fractal power-law behavior and a scale-independent contribution:
 \beq
{\cal L}(\varepsilon) = {\cal L}_0\; \left \{1+ \left( \frac{\lambda}
{\varepsilon} \right) ^{\tau_F}\right\},
\label{eq.5}
\eeq
where $\lambda$ is an integration constant and where $\tau_F=-b=D_F-1$. One easily verifies that the fractal part of this expression agrees with the principle of relativity applied to scales. Indeed, under a  transformation $\varepsilon \to \varepsilon'$, it transforms as ${\cal L}= {\cal L}_0  ({\varepsilon'}/
{\varepsilon}) ^{\tau_F}$ and therefore it depends only on the ratio between scales and not on the individual scales themselves.

This result indicates that, in a general way, fractal functions are the sum of a differentiable part and of a non-differentiable (fractal) part, and that a spontaneous transition is expected to occur between these two behaviors.

On the basis of this elementary solution, many generalisations that may be relevant in biology have been obtained, in particular:\\

\noindent-- log-periodic corrections to power laws:
\begin{equation}
\label{24.}
{\cal L}(\varepsilon )  = a  \, \varepsilon^\nu \, [ 1 + b \cos(\omega  \ln\varepsilon ) ].    
\end{equation}
which is a solution of a second-order differential wave equation in scales.\\

\noindent-- laws of ``scale dynamics" showing a ``scale acceleration":
\begin{equation}
\label{18.}
\tau_F   =  \frac{1}{G}  \ln \left(\frac{\lambda_0 }{\varepsilon }\right), \;\;\;\;\; \ln \left(\frac{{\cal L}}{{\cal L} _{0}}\right)  =  \frac{1}{2G}  \ln ^{2} \left(\frac{\lambda_0 }{\varepsilon }\right).    
\end{equation}
This law may be the manifestation of a constant ``scale force", which describes the difference with the free self-similar case (in analogy with Newton's dynamics of motion). In this case the fractal dimension is no longer constant, but varies in a linear way in terms of the logarithm of resolution. Many manifestations of such a behavior have been identified  in human and physical geography \cite{Forriez2010, Nottale2012C}.\\

\noindent-- laws of special scale relativity \cite{Nottale1992}:
\begin{equation}
\label{14}
 \ln \frac{{\cal L}(\varepsilon )}{{\cal L}_{0}}  =   \frac{\tau_{0 } \, \ln(\lambda_{0}/\varepsilon )}{\sqrt{1 - \ln ^{2}(\lambda  _{0}/\varepsilon ) / \ln ^{2}(\lambda  _{0}/{\lambda_H} )}}    ,   
\end{equation}
\begin{equation}
\label{15}
 \tau_F (\varepsilon )   =   \frac{\tau_0}{\sqrt{1 - \ln ^{2}(\lambda  _{0}/\varepsilon ) / \ln ^{2}(\lambda  _{0}/{\lambda_H} )}}    .    
\end{equation}
This case may not be fully relevant in biology, but we recall it here because it is one of the most profound manifestations of scale relativity. Here the length (i.e. the fractal coordinate) and the `djinn' (variable fractal dimension minus topological dimension) $\tau_F=D_F-1$ have become the components of a vector in scale space. In this new law of scale transformation, a limiting scale appears, $\lambda_H$, which is impassable and invariant under dilations and contractions, independently of the reference scale  $\lambda_0$. We have identified this invariant scale to the Planck length $l_\Pl=\sqrt{\hbar G/c^3}$ toward small scales, and to the cosmic length $\Lu=1/\sqrt{\Lambda}$ (where $\Lambda$ is the cosmological constant) toward large scales \cite{Nottale1992, Nottale1993, Nottale2011}.

Many other scale laws can be constructed as expressions of Euler-Lagrange equations in scale space, which give the general form expected for these laws \cite[Chap. 4]{Nottale2011}.

\subsection{Laws of motion}
The laws of motion in scale relativity are obtained by writing the fundamental equation of dynamics (which is equivalent to a geodesic equation in the absence of an exterior field) in a fractal space. The non-differentiability and the fractality of coordinates implies at least three consequences \cite{Nottale1993, Nottale2011}:

\noindent (1) The number of possible paths is infinite. The description therefore naturally becomes non-deterministic and probabilistic. These virtual paths are identified to the geodesics of the fractal space. The ensemble of these paths constitutes a fluid of geodesics, which is therefore characterized by a velocity field.

\noindent (2) Each of these paths is itself fractal. The velocity field is therefore a fractal function, explicitly dependent on resolutions and divergent when the scale interval tends to zero (this divergence is the manifestation of non-differentibility).

\noindent (3) Moreover, the non-differentiability also implies a two-valuedness of this fractal function, ($V_+, \; V_-$). Indeed, two definitions of the velocity field now exist, which are no longer invariant under a transformation $|dt| \to -|dt|$ in the nondifferentiable case. 

These three properties of motion in a fractal space lead to describing the geodesic velocity field in terms of a complex fractal function $\tilde{\Vsc}= (V_+ + V_-)/2 - i(V_+ - V_-)/2$. The $(+)$ and $(-)$ velocity fields can themselves be decomposed in terms of a differentiable part $v_\pm$ and of a fractal (divergent) fluctuation of zero mean $w_\pm$, i.e., $V_\pm=v_\pm + w_\pm$ and therefore the same is true of the full complex velocity field, $\tilde{\Vsc}={\Vsc}(x,y,z,t)+{\cal W}(x,y,z,t,dt)$. 

Jumping to elementary displacements along these geodesics, this reads 
$dX_{\pm} = d_{\pm}x + d\xi_{\pm}$, with (in the case of a critical fractal dimension $D_F=2$ for the geodesics)
\begin{equation}
d_{\pm} x= v_{\pm} \; dt, \;\;\;
d\xi_{\pm}=\zeta_{\pm} \, \sqrt{2 \cal{D}}  \, |dt|^{1/2}.
\label{eq.20bis}
\end{equation}
This case is particularly relevant since it corresponds to a Markov-like situation of loss of information from one point to the following, without correlation nor anti-correlation.
Here $\zeta_{\pm}$ represents a dimensionless stochastic variable such that $<\!\!\zeta_{\pm}\!\!>=0$ and $<\!\!\zeta_{\pm}^2\!\!>=1$. The parameter $\cal{D}$ characterizes the amplitude of fractal fluctuations.

These various effects can be combined under the construction of a total derivative operator \cite{Nottale1993} :
\begin{equation}
\frac{\widehat{d}}{dt} =  \frac{\partial}{\partial t} + {\cal V}. \nabla -i {\cal  D} \Delta.
\end{equation} 
The fundamental equation of dynamics becomes, in terms of this operator
\beq
m \, \frac{\dfr}{dt} \,{\cal V} =- \nabla \phi.
\label{eq.37}
\eeq
In the absence of an exterior field $\phi$, this is a geodesic equation (i.e., a free inertial Galilean-type equation). 

The next step consists of making a change of variable in which one connects the velocity field $\Vsc=V-iU$ to a function $\psi$ according to the relation
\beq
{\cal V} = - i \, \frac{S_0}{m} \, \nabla \ln \psi.
\label{eq.41}
\eeq
The parameter $S_0$ is a constant for the system considered (it identifies to the Planck constant $\hbar$ in standard quantum mechanics). Thanks to this change of variable, the equation of motion can be integrated under the form of a Schr\"odinger equation \cite{Nottale1993, Nottale2011} generalized to a constant different from $\hbar$,
\beq
{\cal D}^2 \Delta \psi + i {\cal D} \frac{\partial}{\partial t} \psi -
 \frac{\phi}{2m}\psi = 0,
\label{eq.54}
\eeq
where the two parameter introduced hereabove, $S_0$ and $\cal D$, are linked by the relation:
\beq
S_0=2 m \Ds.
\label{Compton}
\eeq
In the case of standard quantum mechanics, $S_0= \hbar$, so that $\Ds$ is a generalisation of the Compton length (up to the constant $c$) and Eq.~(\ref{Compton}) is a generalisation of the Compton relation
\beq
\lambda_C=\frac{2 \Ds}{c}= \frac{\hbar}{mc}.
\eeq
We obtain the same result by using the full velocity field including the fractal fluctuations of zero mean \cite{Nottale2011}. This implies the possible existence of fractal solutions for quantum mechanical equations \cite{Berry1996}. 

By setting finally $\psi = \sqrt{P} \times e^{i \theta}$, with $V=2 \Ds \nabla \theta$, one can show (see \cite{Nottale2007C, Nottale2011} and next section) that $P= |\psi|^2$ gives the number density of virtual geodesics. This function becomes naturally a density of probability or a density of matter or of radiation according to the various conditions of an actual experiment (one particle, many particles or a radiation flow). The function $\psi$, being solution of the Schr\"odinger equation and subjected to the Born postulate and to the Compton relation, owns therefore most of the properties of a wave function.

Reversely, the density $\rho$ and the velocity field $V$ of a fluid in potential motion can be combined in terms of a complex function $\psi= \sqrt{\rho} \times e^{i \theta}$ which may become a wave function solution of a Schr\"odinger equation under some conditions, in particular in the presence of a quantum-type potential (see next section).

 \section{Multiple representations}

After this brief summary of the theory (see more details in \cite{Nottale2011}), let us now consider some of its aspects that may be particularly relevant to applications in biology. One of them is the multiplicity of equivalent representations of the same equations. Usually, classical deterministic equations, quantum equations, stochastic equations, fluid mechanics equations, etc.. correspond to different systems and even to different physical laws. But in the scale relativity framework, they are unified as being different representations of the same fundamental equation (the geodesic equation of relativity), subjected to various changes of variable. This is a particularly useful tool in biophysics, which makes often use of diffusion equations of the Fokker-Planck type or of fluid mechanics equations.

\subsection{Geodesic representation}

The first representation, which can be considered as the root representation, is the geodesic one. The two-valuedness of the velocity field is expressed in this case in terms of the complex velocity field $\Vsc=V-i U$. It implements what makes the essence of the principle of relativity, i.e., the equation of motion must express the fact that any motion should disappear in the proper system of coordinates:
\beq
\Vsc=0.
\eeq
By deriving this equation with respect to time, it takes the form of a free inertial equation devoid of any force:
\beq
\frac{\dfr }{dt}\Vsc=0,
\eeq
where the ``covariant" derivative operator ${\dfr}/{dt}$ includes the terms which account for the effects of the geometry of space-(time). In the case of a fractal space, it reads as we have seen
\beq
\frac{\dfr}{\d t}=\frac{\d}{\d t} + {\cal V} . \nabla -i {\cal D} \Delta.
\eeq

\subsection{Quantum-type representation}
We have recalled in the previous section how a wave function $\psi$ can be introduced from the velocity field of geodesics:
\beq
\Vsc= -2 i \Ds \, \nabla  \ln \psi.
\eeq
This mean that the doubling of the velocity field issued from non-diifferentiability is expressed in this case in terms of the modulus and the phase of this wave function. This allows integration of the equation of motion in the form of a Schr\"odinger equation,
\beq
 {\cal D}^2 \Delta \psi + i  \, {\cal D}\frac{{\d}}{\d t}  \psi - \frac{\phi}{2 m} \,\psi = 0.
 \label{Schro}
\eeq
By making explicit the modulus and the phase of the wave function, $\psi = \sqrt{P} \times e^{i \theta}$, where the phase is related to the classical velocity field by the relation  $V=2\Ds \, \nabla \theta $, one can give this equation the form of hydrodynamics equations including a quantum potential. Moreover, it has been recently shown that this transformation is reversible, i.e., by adding a quantum-like potential energy to a classical fluid, it becomes described by a Schr\"odinger equation and therefore acquires some quantum-type properties \cite{Nottale2009, Nottale2011}.

\subsection{Fluid representation with macroquantum potental}

It is also possible, as we shall now see, to go directly from the geodesic representation to the fluid representation without writing the Schr\"odinger equation.

To this purpose, let us express the complex velocity field in terms of the classical (real) velocity field $V$ and of the number density of geodesics $P_N$, which is equivalent as we have seen hereabove to a probability density $P$:
\beq
\Vsc= V- i \Ds \nabla \ln P.
\eeq
The quantum covariant derivative operator thus reads
\beq
\frac{\dfr}{\d t}=\frac{\d}{\d t} + V. \nabla -i \Ds \: ( \nabla \ln P.\nabla+\Delta).
\eeq
The fundamental equation of dynamics becomes (introducing also an exterior scalar potential $\phi$):
\beq
\l( \frac{\d}{\d t} + V. \nabla -i \Ds \: ( \nabla \ln P.\nabla+\Delta) \r) ( V- i \Ds \nabla \ln P)=-\frac{ \nabla \phi}{m}.
\eeq
The imaginary part of this equation,
\beq
 \Ds \: \l\{  ( \nabla \ln P.\nabla+\Delta)  V + \l( \frac{\d}{\d t} + V. \nabla\r)\nabla \ln P \r\}=0,
\eeq
takes, after some calculations, the following form
\beq
\nabla\l\{   \frac{1}{P}  \l( \frac{\d}{\d t} + \text{div} ( P V) \r) \r\}=0,
\eeq
and it can finally be integrated in terms of a continuity equation
\beq
\frac{\d P}{\d t} + \text{div} ( P V)=0.
\eeq
The real part,
\beq
\l( \frac{\d}{\d t} + V. \nabla\r) V=-\frac{ \nabla \phi}{m} +\Ds^2 \: ( \nabla \ln P.\nabla+\Delta) \nabla \ln P,
\eeq
takes the form of an Euler equation, 
\beq
m \l( \frac{\d}{\d t} + V. \nabla\r) V=- \nabla \phi +2 m \Ds^2 \:  \nabla \l( \frac{ \Delta \sqrt{P}}{\sqrt{P}} \r),
\eeq
and it therefore describes a fluid subjected to an additional quantum-type potential 
\beq
Q=-2 m \Ds^2 \: \frac{ \Delta \sqrt{P}}{\sqrt{P}}.
\eeq
It is remarkable that we have obtained this result directly, without passing through a quantum-type representation using a wave function nor through a Schr\"odinger equation.

The additional ``fractal" potential is obtained here as a mere manifestation of the fractal geometry of space, in analogy with Newton's potential emerging as a manifestation of the curved geometry of space-time in Einstein's relativistic theory of gravitation. We have suggested (\cite{Nottale2011} and references therein) that this geometric energy could contribute to the effects which have been attributed in astrophysics to a missing ``dark matter" (knowing that all attempts to directly observe this missing mass have so far failed). Another suggestion, relevant to biology, is that such a potential energy could play an important role in the self-organisation and in the morphogenesis of living systems \cite{Nottale2001, Nottale2008}.

\subsection{Coupled two-fluids}
Another equivalent possible representation consists of separating the real and imaginary parts of the complex velocity field, 
\beq
\Vsc= V- i U.
\eeq
One obtains in this case a system of equations that describe the velocity fields of two fluids strongly coupled together,
\beq
 \l( \frac{\d}{\d t} + V. \nabla \r) V=(U. \nabla+ \Ds  \Delta ) U - \nabla \l( \frac{\phi}{m}\r),
\eeq
\beq
 \l( \frac{\d}{\d t} + V. \nabla \r) U=-(U. \nabla+ \Ds  \Delta ) V.
\eeq
This representation may be useful in, e.g., numerical simulations of scale relativity / quantum processes \cite{Hermann1997}.

\subsection{Diffusion-type representation}
The fundamental two-valuedness which is a consequence of nondifferentiability has been initially described in terms of two mean velocity fields $v_+$ and $v_-$, which transform one into the other by the reflexion $|dt| \leftrightarrow -|dt|$.  It is therefore possible to write the equations of motion directly in terms of these two velocity fields. The representation obtained in this way implements the diffusive character of a fractal space and is therefore particularly interesting for biophysical applications. Indeed, one obtains the standard Fokker-Planck equation for the velocity $v_+$, as for a classical stochastic process :
\beq
\frac{\d P} {\d t} + \text{div} (P v_+)= \Ds \Delta P,
\label{FP}
\eeq
where the parameter $\Ds$ plays the role of a diffusion coefficient. On the contrary, the equation obtained for the velocity field $v_-$ does not correspond to any classical process:
\beq
\frac{\d P} {\d t} + \text{div} (P v_-)= -\Ds \Delta P.
\eeq
This equation is derived from the geodesic equation on the basis of nondifferentiability, but it cannot be set as a founding equation in the framework of a standard diffusion process as was proposed by Nelson \cite{Nelson1966}, since it becomes self-contradictory with the backward Kolmogorov equation generated by such a classical process \cite{Grabert1979, Wang1993, Nottale1997A} \cite[p.~384]{Nottale2011}.

\subsection{A new form of quantum-type potential}
However, one may remark that the previous representation is not fully coherent, since it involves three quantities $P$, $v_+$ and $v_-$ instead of two expected from the velocity doubling. Therefore it should be possible to obtain a system of equations involving only the probability density $P$ and one of the velocity fields, here $v_+$. To this purpose, one remarks that $v_-$ is given in terms of these two quantities by the relation:
\beq
v_-= v_+ -2 \Ds \nabla \ln P.
\eeq
We also recall that 
\beq
V= v_+    - \Ds \nabla \ln P.
\label{vvv}
\eeq
The energy equation now reads
\beq
E= \frac{1}{2} m \,V^2 + Q + \phi=  \frac{1}{2} m  \,(v_+ - \Ds \nabla \ln P)^2 + Q + \phi,
\eeq
where the macroquantum potential can be written
\beq
Q=-2 m \Ds^2 \: \frac{ \Delta \sqrt{P}}{\sqrt{P}}=- m \Ds^2 \: \l\{ \Delta \ln P + \frac{1}{2} (\nabla \ln P)^2 \r\}.
\eeq
One of the terms of this ``fractal potential" is therefore compensated while another term appears, so that we obtain:
\beq
E= \frac{1}{2} m \,v_+^2 +  \phi -m\, \Ds \,v_+. \nabla \ln P -m \Ds^2 \Delta \ln P.
\eeq
We finally obtain a new representation in terms of a Fokker-Planck equation, which contains the diffusive term $ \Ds \Delta P$ in addition to the continuity equation obtained in the case of the fluid representation ($V$, $P$), and an energy equation which includes a new form of quantum potential:
\beq
\frac{\d P} {\d t} + \text{div} (P v_+)= \Ds \Delta P,
\eeq
\beq
E= \frac{1}{2} m \,v_+^2 +  \phi + Q_+,
\eeq
where the new quantum-type potential reads
\beq
Q_+=- m\, \Ds \, (v_+ . \nabla \ln P+ \Ds \Delta \ln P).
\eeq
It now depends not only on the probability density $P$, but also on the velocity field $v_+$.

This derivation is once again reversible. This means that a classical diffusive system described by a standard Fokker-Planck equation which would be subjected to such a generalized quantum-type potential would be spontaneously transformed into a quantum-like system described by a Schr\"odinger equation (\ref{Schro}) acting on a wave function
 $\psi= \sqrt{P} \times e^{i \, \theta}$ where  $V=2\Ds \nabla \theta$. Thanks to Eq.~(\ref{vvv}), this wave function is defined in terms of  $P$ and $v_+$ as
\beq
\psi= \sqrt{P}^{1-i} \times e^{i \, \theta_+},
\eeq
where $v_+= 2 \Ds \, \nabla \theta_+$. 

Such a system, although it is initially diffusive, would therefore acquire some quantum-type properties, but evidently not all of them: the behaviors of coherence, unseparability, indistinguishability or entanglement are specific of a combination of quantum laws and elementarity \cite{Weisskopf1989} and cannot be recovered in such a context. 

This is nevertheless a remarkable result, which means that a partial reversal of diffusion and a transformation of a classical diffusive system into a quantum-type self-organized one should be possible by applying a quantum-like force to this system. This is possible in an actual experiment consisting of a retro-active loop involving continuous measurements, not only of the density \cite{Nottale2009} but also of the velocity field $v_+$, followed by a real time application on the system of a classical force $F_{Q+}=-\nabla Q_+$ simulating the new macroquantum force \cite{Nottale2012}. 

One may also wonder whether living systems, which already work in terms of such a feedback loop (involving sensors, then cognitive processes, then actuators) could have naturally included such kinds of quantum-like potentials in their operation through the selection / evolution process, simply because it provides an enormous evolutionary advantage due to its self-organization and morphogenesis negentropic capabilities \cite{Nottale2008} \cite[Chap.~14]{Nottale2011}.

\subsection{Quantum potential reversal}
One of the recently obtained results which may be particularly relevant to the understanding of living systems concerns the reversal of the quantum-type potentiel. What happens when the potential energy keeps exactly the same form, as given by $ \Delta \sqrt{P}/{\sqrt{P}}$ for a given distribution $P(x,y,z)$, while its sign is reversed ? In other words, to what kind of process does the equation 
\beq
 \l( \frac{\d}{\d t} + V. \nabla\r) V=- \frac{ \nabla \phi}{m} -2  \Ds^2 \:  \nabla\frac{ \Delta \sqrt{P}}{\sqrt{P}},
\eeq
correspond ? 

We have shown \cite{Nottale2008, Nottale2011} that such an Euler equation, when it is combined with a continuity equation, can no longer be integrated under the form of a generalized Schr\"odinger equation. This process is therefore no longer self-organizing. On the contrary, this is a classical diffusive process, characterized by an entropy increase proportional to time. 

Indeed, let us start from a Fokker-Planck equation
\beq
\frac{\d P} {\d t} + \text{div} (P v)= D \Delta P,
\eeq
which describes a classical diffusion process with diffusion coefficient $D$. Then make the change of variable
\beq
V=v- D \nabla \ln P.
\eeq
One finds after some calculations that $V$ and $P$ are now solutions of a continuity equation
\beq
\frac{\d P} {\d t} + \text{div} (P v)=0,
\eeq
and of an Euler equation which reads
\beq
 \l( \frac{\d}{\d t} + V. \nabla\r) V=-2 D^2 \:  \nabla\frac{ \Delta \sqrt{P}}{\sqrt{P}}.
\eeq
In other words, we have obtained an hydrodynamical description of a standard diffusion process in terms of a ``diffusion potential" which is exactly the reverse of the macroquantum potential. 

We have suggested that this behavior may be relevant for the understanding of cancer \cite{Nottale2008, Nottale2011} (see also \cite{Waliszewski2001} about the relationship between fractal geometry and tumors), since a mere change of sign of the additional potential leads to dramatic consequences: the self-organizing, morphogenetic and structuring character of the system is instantaneouly changed to a diffusive, anti-structuring desorganization.

\section{Quantum potentials in high-temperature superconductivity}

\subsection{Ginzburg-Landau non-linear Schr\"odinger equation}

The phenomenon of superconductivity is one of the most fascinating of physics. It lies at the heart of a large part of modern physics. Indeed, besides its proper interest for the understanding of condensed mater, it has been used as model for the construction of the electroweak theory through the Higgs field and of other theories in particle physics and in other sciences. 
	
Moreover, superconductivity (SC) has led physicists to deep insights about the nature of matter. It has shown that the ancient view of matter as something ``solid", in other words ``material", was incorrect. The question: ``is it possible to walk through walls" is now asked in a different way. Nowadays we know that it is not a property of matter by itself which provides it qualities such as solidity or ability to be crossed, but its interactions. 

A first relation of SC with the scale relativity approach can be found in its phenomenological Ginzburg-Landau equation. Indeed, one can recover such a non-linear Schr\"odinger equation simply by adding a quantum-like potential energy to a standard fluid including a pressure term \cite{Nottale2009}. 

Consider indeed an Euler equation with a pressure term and a quantum potential term: 
\begin{equation}
\label{AAA1}
 \l(\frac{\partial}{\partial t} + V \cdot \nabla\r) V  = -\nabla \phi-\frac{\nabla p}{\rho}+2{\cal D}^2 \,\nabla \l( \frac{\Delta \sqrt{\rho}}{\sqrt{\rho}}\r).
\end{equation}
When ${\nabla p}/{\rho}=\nabla w$ is itself a gradient, which is the case of an isentropic fluid, and, more generally, of every cases when there is a state equation which links $p$ and $\rho$, its combination with the continuity equation can be still integrated in terms of a Schr\"odinger-type equation \cite{Nottale1997A},
\beq
{\cal D}^2 \Delta \psi + i {\cal D} \frac{\partial}{\partial t} \psi - \frac{\phi+w}{2}\psi = 0.
\eeq
In the sound approximation, the link between pressure and density writes $p-p_0=c_s^2(\rho-\rho_0)$, where $c_s$ is the sound speed in the fluid, so that $\nabla p/\rho=c_s^2 \, \nabla \ln \rho$. Moreover, when $\rho-\rho_0 \ll \rho_0$, one may use the additional approximation $c_s^2 \, \nabla \ln \rho \approx (c_s^2 /\rho_0) \nabla \rho$, and the equation obtained takes the form of the  Ginzburg-Landau equation of superconductivity \cite{Landau9},
\beq
{\cal D}^2 \Delta \psi + i {\cal D} \frac{\partial}{\partial t} \psi - \beta \, |\psi|^2 \, \psi= \frac{1}{2} \, \phi  \; \psi,
\eeq
with $\beta={c_s^2}/{2 \rho_0}$. In the highly compressible case, the dominant pressure term is rather of the form  $p \propto \rho^2$, so that $p/\rho \propto \rho= |\psi|^2$, and one still obtains a non-linear Schr\"odinger equation of the same kind \cite{Nore1994}.

The intervention of pressure is highly probable in living systems, so that such an equation is expected to be relevant in theoretical systems biology. Laboratory experiments aiming at implementing this transformation of a classical fluid into a macroscopic quantum-type fluid are presently under development \cite{Nottale2009C, Nottale2012}.

\subsection{A quantum potential as origin of Cooper pairs in HTS ?}
Another important question concerning SC is that of the microscopic theory which gives rise to such a macroscopic phenomenological behavior. 

In superconducting materials, the bounding of electrons in Cooper pairs transforms the electronic gas from a fermionic to a bosonic quantum fluid. The interaction of this fluid with the atoms of the SC material becomes so small that the conducting electrons do not ``see" any longer the material. The SC electrons become almost free, all resistance is abolished and one passes from simple conduction to superconduction. 

In normal superconductors, the pairing of electrons is a result of their interaction with phonons (see, e.g., \cite{deGennes1989}). But since 1985, a new form of superconductivity has been discovered which has been named ``high temperature superconductivity" (HTS) because the critical temperature, which was of the order of a few kelvins for normal SC, has reached up to 135 K. However, though it has been shown that HTS is still due to the formation of Cooper pairs, the origin of the force that pairs the electrons can no longer be phonons and still remains unknown. Actually, it can be proved that any attractive force between the electrons, as small it could be, would produce their Cooper pairing \cite{Landau5}.

Therefore the problem of HTS can be traced back to that of identifying the force that links the electrons. We have suggested that this force actually derives from a quantum potential \cite{Nottale2013}.

Most HTS are copper oxide compounds in which superconductivity arises when they are doped either by extra charges but more often by `holes' (positive charge carrier).  
Moreover, a systematic electronic inhomogeneity has been reported at the microscopic level, in particular in compounds like Bi$_2$Sr$_2$CaCu$_2$O$_{8+x}$ \cite {Pan2001}, the local density of states (LDOS) showing `hills' and `valley' of size $\sim$ 30 Angstroms, strongly correlated with the SC gap. Actually, the minima of LDOS modulations preferentially occur at the dopant defects \cite{McElroy2005}. The regions with sharp coherence peaks, usually associated with strong superconductivity, are found to occur between the dopant defect clusters, near which the SC coherence peaks are suppressed.

Basing ourselves on these observations, we have suggested that, at least in this type of compound, the electrons can be traped in the quantum potential well created by these electronic modulations.

Let us give here a summary of this new proposal. We denote by $\psi_n$ the wave function of doping charges which have diffused from the initial site of dopant defects, and by $\psi_s$ the wave function of the fraction of carriers which will be tied in Cooper pairs (only 19-23 \% of the total doping induced charge joins the superfluid near optimum doping).

We set  $\psi_n= \psi_s +  \psi_d$, where $ \psi_d$ is the wave function of the fraction of charges which do not participate in the superconductivity.

The doping induced charges constitutes a quantum fluid which is expected to be the solution of a Schr\"odinger equation (here of standard QM, i.e. written in terms of the microscopic Planck's constant $\hbar$)
\beq
\frac{\hbar^2}{2m} \Delta \psi_n + i \hbar \frac{ \d \psi_n}{\d t}= \phi\;  \psi_n,
\eeq 
where $\phi$ is a possible external scalar potential, and where we have neglected the magnetic effects as a first step.

Let us separate the two contributions  $\psi_s$ and  $\psi_d$ in this equation. We obtain:
\beq
\frac{\hbar^2}{2m} \Delta \psi_s + i \hbar \frac{ \d \psi_s}{\d t}- \phi\;  \psi_s =-\frac{\hbar^2}{2m} \Delta \psi_d - i \hbar \frac{ \d \psi_d}{\d t}+\phi \; \psi_d.
\eeq
We can now introduce explicitly the probability densities $n$ and the phases $\theta$ of the wave functions  $\psi_s= \sqrt{n_s} \times e^{i \theta_s}$ and $\psi_d= \sqrt{n_d} \times e^{i \theta_d}$. The velocity fields of the (s) and (d) quantum fluids are given by $V_s=(\hbar/m) \nabla \theta_s$ and $V_d=(\hbar/m) \nabla \theta_d$. As we have seen hereabove, a Schr\"odinger equation 
can be put into the form of fluid mechanics-like equations, its imaginary part becoming a continuity equation and the derivative of its real part becoming a Euler equation with quantum potential. Therefore the above equation can be written as:
\beq
\frac{\d V_s}{\d t} + V_s. \nabla V_s =- \frac{\nabla \phi}{m} -\frac{\nabla Q_s}{m}-\l(\frac{\d V_d}{\d t} + V_d. \nabla V_d + \frac{\nabla Q_d}{m}\r)
\eeq
\beq
\frac{\d n_s }{ \d t} + \text{div}(n_s V_s)=-\frac{\d n_d }{ \d t} - \text{div}(n_d V_d).
\eeq
But the (d) part of the quantum fluid, which is not involved in the superconductivity, remains essentially static, so that $V_d=0$ and $\d n_d /{ \d t}=0$. Therefore we obtain for the quantum fluid (s) a new system of fluid equations:
\beq
\frac{\d V_s}{\d t} + V_s. \nabla V_s =- \frac{\nabla \phi}{m} -\frac{\nabla Q_s}{m}- \frac{\nabla Q_d}{m},
\eeq
\beq
\frac{\d n_s }{ \d t} + \text{div}(n_s V_s)=0,
\eeq
which can be re-integrated under the form of a Schr\"odinger equation
\beq
\frac{\hbar^2}{2m} \Delta \psi_s + i \hbar \frac{ \d \psi_s}{\d t}- (\phi + Q_d)  \psi_s =0.
\eeq
It therefore describes the motion of electrons (s), represented by their wave function $\psi_s$, in a potential well given by the exterior potential $\phi$, but also by an interior quantum potential $Q_d$ which just depends on the local fluctuations of the density $n_d$ of charges,
 \beq
 Q_d= -\frac{\hbar^2}{2m} \frac{\Delta \sqrt{n_d}}{ \sqrt{n_d}}.
 \eeq
 
Even if in its details this rough model is probably incomplete, we hope this proposal, according to which the quantum potential created by the dopants provides the attractive force needed to link electrons into Cooper pairs, to be globally correct, at least for some of the existing HT superconductors.

 Many (up to now) poorly understood features of cuprate HTS can be explained by this model. For example, the quantum potential well involves bound states in which two electrons can be trapped with zero total spin and momentum. One can show that the optimal configuration for obtaining bound states is with 4 dopant defects (oxygen atoms), which bring 8 additional charges. One therefore expects a ratio $n_s/n_n= 2/(8+2)=0.2$ at optimal doping. This is precisely the observed value \cite{Tanner1998}, for which, to our knowledge, no explanation existed up to now. 

The characteristic size of LDOS wells of $\sim$ 30 Angstroms is also easily recovered in this context: the optimal doping being $p=0.155=1/6.5$, the 8 to 10 charges  present in the potential well correspond to a surface $(8-10) \times 6.5= (52-65) =(7.2-8.1)^2$ in units of $d_{\rm CuO}=3.9$ Angstroms, i.e. 28-32 Angstroms as observed experimentally.

In this context, the high critical temperature superconductivity would be a geometric multiscale effect. In normal SC, the various elements which permit the superconductivity, Cooper pairing of electrons, formation of a quantum bosonic fluid and coherence of this fluid are simultaneous. In HTS, under the quantum potential hypothesis, these elements would be partly disconnected and related to different structures at different scales (in relation to the connectivity of the potential wells), achieving a multi-scale fractal structure \cite{Fratini2010}.

If confirmed, this would be a nice application of the concept of quantum potentials \cite{Bohm}, here in the context of standard microscopic quantum mechanics.

\section{Scale relativity in nondifferentiable velocity-space}

\subsection{Analogy between turbulence and living systems}

Living systems are well known to exhibit fractal structures from very small scales up to the organism size and even to the size of the collective entities (e.g., a forest made of trees). Therefore it is relevant to assess and quantify these properties with sophisticated models. 

Some advanced fractal and multifractal models have been developed in the field of turbulence because fractals are the basic fundamental feature of chaotic fluid dynamics \cite{Frisch1982}. They have been described since the famous law of Kolmogorov, known as K41 \cite{Kolmogorov1941}. In the atmosphere, scale laws are observed from micrometers up to thousands of kilometers. Turbulence can be described as flow of energy injected at large scale that cascades into smaller and smaller structures. This process redirects the energy into all directions and it is ultimately dissipated into heat at the smallest scale. 

There is therefore a strong analogy with living systems. An investigation of turbulence versus living systems is particularly interesting as there are a number of common points:\\
- Dissipation: both turbulent flows and living systems are dissipative. \\
- Non-isolated: existence of source and sink of energy.\\
- Out of equilibrium. \\
- Chaotic. \\
- Existence of stationary structures.  Individual "particles" enter and go out in a very complex way, while the overall structure grows (growth of living systems, development of turbulence) then remains stable on a long time scale.\\
- Fundamentally multi-scale and multi-fractal structuring. \\
- Injection of energy at an extreme scale with dissipation at the other (the direction of the multiplicative cascade is reversed in living systems compared to laboratory turbulence).\\
etc..

\subsection{Application of scale relativity to turbulence}

In a recent work, L. de Montera has suggested an original application of the scale relativity theory to the yet unsolved problem of turbulence in fluid mechanics \cite{deMontera2013}. He has remarked that the Kolmogorov scaling of velocity increments in a Lagrangian description (where one follows an element of fluid, for example thanks to a seeded micro particle \cite{LaPorta2001}),
\beq
\delta v \propto {|\delta t|}^{1/2},
\label{K41}
\eeq
was exactly similar to the fractal fluctuation Eq.~(\ref{eq.20bis}) which is at the basis of the scale relativity description. 

The difference is that coordinates remain differentiable, while in this new context velocity becomes non-differentiable, so that accelerations $a= \delta v / \delta t \propto {|\delta t|}^{-1/2}$ become scale-divergent. Although this power law divergence is clearly limited by the dissipative Kolmogorov small scale, it is nevertheless fairly supported by experimental data, since acceleration of up to 1500 times the acceleration of gravity have been measured in turbulent flows \cite{LaPorta2001,Voth2008}).

De Montera's suggestion therefore amounts to apply the scale relativity method after an additional order of differentiation of the equations. The need for such a shift has already been remarked in the framework of stochastic models of turbulence \cite{Beck2005,Sawford1991}.

Let us consider here some possible implications of this new proposal.

The necessary conditions which underlie the construction of the scale relativity covariant derivative are very clearly fulfilled for turbulence (now in velocity space): 

\noindent (1) The chaotic motion of fluid particles implies an infinity of possible paths.

\noindent (2) Each of the paths (realisations of which are achieved by test particles of size $<100\: \mu$m in a lagrangian approach, \cite{Voth2008}) are of fractal dimension $D_F=2$ in velocity space, at least in the K41 regime (Eq.~\ref{K41}). 

\noindent (3) The two-valuedness of acceleration is manifest in turbulence data. As remarked by Falkovich et al. \cite{Falkovich2012}, the usual statistical tools of description of turbulence 
(correlation function, second order structure function, etc...)
are reversible, while turbulence, being a dissipative process, is fundamentally irreversible. The two-valuedness of derivative is just a way to account for the symmetry breaking under the time scale reflexion $\delta t  \to - \delta t$. Among the various ways to describe this doubling \cite{Nottale2012B}, one of them is particularly adapted to comparison with turbulence data. It consists of remarking that the calculation of a derivative involves a Taylor expansion
\beq
\frac{dX}{dt}= \frac{X(t+dt)-X(t)}{dt}=\frac{(X(t)+X'(t) dt + \frac{1}{2} X''(t) dt^2 +...)-X(t)}{dt},\eeq
so that one obtains
\beq
\frac{dX}{dt}=X'(t)  + \frac{1}{2} X''(t) \, dt + ...
\eeq
For a standard non fractal function, the contribution $ \frac{1}{2} X''(t) dt $ and all the following terms of higher order vanish when $dt \to 0$, so that one recovers the usual result $dX/dt=X'(t)$. But for a fractal function such that its second derivative is scale divergent as  $X''(t) \propto 1/dt$, the second order term can no longer be neglected and must contribute to the definition of the derivative \cite[Sec.~3.1]{Nottale2011}. Therefore one may write
\beq
\frac{d_+X}{dt}=X'(t)  + \frac{1}{2} X''(t) \; |dt| , \;\;\; \frac{d_-X}{dt}=X'(t)  - \frac{1}{2} X''(t)\; |dt|,
\eeq
then
\beq
\frac{\dfr X}{dt}= \frac{d_+ + d_-}{2 dt}X - i \, \frac{d_+ - d_-}{2 dt}X= X'(t) - i \, \frac{1}{2} X''(t) \;|dt|.
\eeq
Lagrangian measurements of turbulence data \cite{Mordant2001, Mordant2001B} confirm this expectation. One finds that the acceleration $a=v'$ and its increments $da=v'' dt$ are indeed of the same numerical order: in these data, the dispersions are respectively $\sigma_a= 280 \:{\rm m/s}^2$ vs $\sigma_{da}= 220 \:{\rm m/s}^2$. This fundamental result fully supports the acceleration two-valuedness on an experimental basis.

\noindent (4) The dynamics is Newtonian: the equation of dynamics in velocity space is the time derivative of the Navier-Stokes equation, i.e.,
\beq
\frac{da}{dt}=\dot{F}.
\eeq
Langevin-type friction terms may occur in this equation but they do not change the nature of the dynamics. They will simply add a non-linear contribution in the final Schr\"odinger equation.

\noindent (5) The range of scales is large enough for a K41 regime to be established: in von Karman laboratory fully developed turbulence experiments, the ratio between the small dissipative scale and the large (energy injection) scale is larger than 1000 and a K41 regime is actually observed \cite{Mordant2001}.

The application of the scale relativity method is therefore fully supported experimentally in this case. Velocity increments $dV$ can be decomposed into two terms, a classical differentiable one $dv$ and a fractal fluctuation:
\beq
dV=dv + \zeta  \,\sqrt{2 \Ds_v \,  dt},
\eeq
where $<\zeta>=0$ and $<\zeta^2>=1$. One recognizes here the K41 scaling in $dt^{1/2}$.
One introduces a complex acceleration field ${\cal A}= a-i \, (da/2)$ and a total `covariant' derivative
\beq
\frac{\dfr}{dt}= \frac{ \d } {dt} + {\cal A}. \nabla_v-i \,\Ds_v \, \Delta_v
\eeq
and then write a super-dynamics equation
\beq
\frac{\dfr}{dt}{\cal A}=\dot{F}.
\eeq
A wave function $\psi$ acting in velocity space can be constructed from the acceleration field,
\beq
{\cal A}=-2 i \,  \Ds_v \nabla_v \ln \psi
\eeq
and the super-dynamics equation can then be integrated under the form of a Schr\"odinger equation including possible non-linear terms (NLT)
\beq
\Ds_v^2 \,  \Delta_v \psi + i \,  \Ds_v \frac{\d \psi} {\d t}= \frac{\phi}{2} \, \psi +NLT,
\eeq
where $\phi$ is a potential (in velocity space) from which the force $\dot{F}$ or part of this force derives.

By coming back to a fluid representation -- but now in terms of the fluid of potential paths -- using as variables $P(v)= |\psi|^2$ and $a(v)$ (which derives from the phase of the wave function), this equation becomes equivalent to the combination of a Navier-Stokes-like equation written in velocity space and a continuity equation,
\beq
\frac{da}{dt}=\dot{F}+ 2 \Ds_v^2 \, \nabla_v \l( \frac{\Delta_v \sqrt{P}}{\sqrt{P}}\r),
\eeq
\beq
\frac{\d P}{\d t} + {\rm div}_v (P a)=0.
\eeq
Therefore we have recovered the same equation from which we started (time derivative of Navier-Stokes equation) but a new term has energed, namely, a quantum-type force which is the gradient of a quantum-type potential in velocity space. One can now re-integrate this equation, and one thus obtains the initial Navier-Stokes equation (in the uncompressible case $\rho=1$ and with a viscosity coefficient $\nu$):
\beq
\l( \frac{\d} {\d t} + v. \nabla \r) v= -\nabla p + \nu \Delta v +2 \Ds_v^2 \int_0^t  \nabla_v \l( \frac{\Delta_v \sqrt{P}}{\sqrt{P}}\r) dt,
\eeq
but with an additional term which manifests the fractality of the flow in velocity space. The value of $\Ds_v$ is directly given, in the K41 regime, by the parameter which commands the whole process, the energy dissipation rate by unit of mass, $\varepsilon$,
\beq
2 \Ds_v= C_0 \varepsilon,
\eeq
where $C_0$ is Kolmogorov's numerical constant (whose estimations vary from 4 to 9). Concerning the two small scale (dissipative) and large scale (energy injection) transitions, one could include them in a scale varying $\Ds_v$, but a better solution consists of keeping $\Ds_v$ constant, then to include the transitions subsequently to the whole process in a global way.

The intervention of such a missing term in developed turbulence is quite possible and  is even supported by experimental data. Indeed, precise experimental measurements of one of the numerical constants which characterize the universal scaling of turbulent flows, $a_0=\nu^{1/2} \varepsilon^{-3/2} \sigma_a^2 $, has given constant values around $a_0=6$ in the developed turbulence domain $R_\lambda \geq 500$ \cite{Voth2008}. However, in the same time, direct numerical simulations (DNS) of Navier-Stokes equations under the same conditions \cite{Vedula1999, Gotoh2001, Ishihara2007} have systematically given values around $a_0=4$, smaller by a factor 2/3. 

Let us derive the scale relativity prediction of this constant. We indeed expect an additional contribution to the DNS, since they use the standard NS equations and do not include the new quantum potential. 

The considered experiments are van Karman-type flows. The turbulence is generated in a flow of water between counter-rotating disks (with the same opposite rotational velocity) in a cylindrical container \cite{Voth2008}. For such experiments the Lagrangian velocity distribution is given with a good approximation by a Gaussian distribution \cite{Mordant2001, Mordant2001B} centered on $v=0$ (in the laboratory reference system). We can therefore easily calculate the velocity quantum potential. We find for these specific experiments
\beq
Q_v=- \Ds_v^2  \l(\frac{v^2-6 \sigma_v^2}{2 \sigma_v^4}\r),
\eeq
where $\sigma_v^2$ is the velocity variance.
Therefore the quantum-like force reads ${\bf F}_{Qv}=-\nabla_v Q_v=  (\Ds_v^2 /\sigma_v^4)\,{\bf v}$, and the additional term in Navier-Stokes equations finally reads, in a reference system whose origin is the center of the cylinder,
\beq
{\bf F}_{Qx}= \int_0^t {\bf F}_{Qv} dt= \frac{\Ds_v^2 }{\sigma_v^4}\; {\bf x},
\eeq
which is just a repulsive harmonic oscillator force. We therefore expect a new geometric contribution to the acceleration variance:
\beq
\sigma_a^2=(\sigma_a)_{cl}^2+ \frac{\Ds_v^4 }{\sigma_v^8}\;\sigma_x^2.
\eeq
Now the parameter $\Ds_v= C_0 \varepsilon/2$ can also, in the K41 regime, be written in function of $\sigma_v$ and of the Lagrangian integral time scale $T_L$ as $\Ds_v={\sigma_v^2}/{T_L}$, while we can take $\sigma_x \approx L$, the Lagrangian length scale, and we obtain the simple expression
\beq
\frac{(\sigma_a)_{cl}^2}{\sigma_a^2}=1- \frac{L^2}{\sigma_a^2 T_L^4}.
\eeq
This ratio has been observed to be $\approx 2/3$ by Voth et al. \cite{Voth2008} (taking the DNS values for $(\sigma_a)_{cl}$ and the experimental ones for $\sigma_a$). The experimental values of $L$, $\sigma_a$ \cite{Voth2008} and of $T_L$ (fitted from the published data) for the same experiments precisely yield values around 2/3, a very satisfactory agreement between the theoretical expectation and the experimental result. 

For example, in one of the experiments with $R_\lambda=690$, Voth et al. have measured $\sigma_a=87$ m/s$^2$ and $L= 0.071$ m \cite{Voth2008}, while the fitted Lagrangian time scale is found to be $T_L=39$ ms, so that ${L}/(\sigma_a T_L^2)=0.54$. For the same experiment, $(a_0)_{\rm DNS}=4.5$ while $(a_0)_{\rm exp}=6.2$, so that $(1-(a_0)_{\rm DNS}/(a_0)_{\rm exp})^{1/2}=0.52$, very close to the theoretical expectation from the scale relativity correction.

Although this is not yet a definitive proof of a quantum-like regime in velocity space for developed turbulence (which we shall search in a finer analysis of turbulence data \cite{Nottale2013}), this adequation is nevertheless a very encouraging result in favor of de Montera's proposal. 

\section{Conclusion}

The theory of scale relativity, thanks to its account, at a profound level, of the fractal geometry of a system, is particularly adapted to the construction and development of a theoretical biology. In its framework, the description of living systems is no longer strictly deterministic. It supports the use of statistical and probabilistic tools in biology, for example as concerns the expression of genes \cite{Laforge2005,Kupiec2008}. 

But it also suggests to go beyond ordinary probabilities, since the description tool becomes a quantum-like (macroscopic) wave function, which is the solution of a generalized Schr\"odinger equation. This involves a probability density such that $P= |\psi|^2$, but also phases, which are built from the velocity field of potential trajectories and yield possible interferences.

Such a Schr\"odinger (or non-linear Schr\"odinger) form of motion equations can be obtained in at least two ways. One way is through the fractality of the biological medium, which is now validated at several scales of living systems, for example in cell walls \cite{Turner2011}. Another way is through the emergence of macroscopic quantum-type potentials, which could be an advantageous character acquired from evolution and selection.

In this framework, one therefore expects a fundamentally wave-like and often quantized character of numerous processes implemented in living systems. In the present contribution, we have concentrated ourselves only on the purely theoretical aspect of the scale relativity approach. But many explicit applications to real biological organisms have also been obtained with success (see \cite{Nottale2000B, Nottale2009B}, \cite[Chap.~14]{Nottale2011} and references therein). 

Several properties that are considered to be specific of biological systems, such as self-organization, morphogenesis, ability to duplicate, reproduce and branch, confinement and multi-scale structuration and integration are naturally obtained in such an approach \cite{Auffray2008, Nottale2008}.  Reversely, the implementation of this type of process in new technological devices involving intelligent feedback loops and quantum-type potentials  could also lead to the emergence of a new form of `artificial life'.\\

{\bf Acknowledgement.} The author gratefully thanks Dr. Philip Turner for his careful reading of the manuscript and for his useful comments.



\end{document}